\documentclass[sigconf]{acmart}

\AtBeginDocument{%
  }

\setcopyright{acmcopyright}
\copyrightyear{2018}
\acmYear{2018}
\acmDOI{XXXXXXX.XXXXXXX}

\acmConference[Conference acronym 'XX]{Make sure to enter the correct
  conference title from your rights confirmation emai}{
  2022}{}
\acmPrice{15.00}
\acmISBN{978-1-4503-XXXX-X/18/06}

\begin{document}

%%
%% The "title" command has an optional parameter,
%% allowing the author to define a "short title" to be used in page headers.
% \title{Vulnerability Prioritization from Hackers' Point of View}
\title{Vulnerability Prioritization: An Offensive Security Approach}
% \subtitle{Work-in-Progress}

%%
%% The "author" command and its associated commands are used to define
%% the authors and their affiliations.
%% Of note is the shared affiliation of the first two authors, and the
%% "authornote" and "authornotemark" commands
%% used to denote shared contribution to the research.
\author{Muhammed Fatih Bulut}
\email{mfbulut@us.ibm.com}
%\orcid{1234-5678-9012}
\affiliation{%
  \institution{IBM T.J. Watson Research Center}
  \city{Yorktown Heights}
  \state{New York}
  \country{USA}
}

\author{Abdulhamid Adebayo}
\email{hamid.adebayo@ibm.com}
%\orcid{1234-5678-9012}
\affiliation{%
  \institution{IBM T.J. Watson Research Center}
  \city{Yorktown Heights}
  \state{New York}
  \country{USA}
}

\author{Daby Sow}
\email{sowdaby@us.ibm.com}
%\orcid{1234-5678-9012}
\affiliation{%
  \institution{IBM T.J. Watson Research Center}
  \city{Yorktown Heights}
  \state{New York}
  \country{USA}
}

\author{Steve Ocepek}
\email{steve.ocepek@ibm.com}
%\orcid{1234-5678-9012}
\affiliation{%
  \institution{IBM Security}
  \city{Armonk}
  \state{New York}
  \country{USA}
}

%%
%% By default, the full list of authors will be used in the page
%% headers. Often, this list is too long, and will overlap
%% other information printed in the page headers. This command allows
%% the author to define a more concise list
%% of authors' names for this purpose.
\renewcommand{\shortauthors}{Bulut et al.}

%%
%% The abstract is a short summary of the work to be presented in the
%% article.

\begin{abstract}

Organizations struggle to handle sheer number of vulnerabilities in their cloud environments. The de facto methodology used for prioritizing vulnerabilities is to use Common Vulnerability Scoring System (CVSS). However, CVSS has inherent limitations that makes it not ideal for prioritization. In this work, we propose a new way of prioritizing vulnerabilities. Our approach is inspired by how offensive security practitioners perform penetration testing. We evaluate our approach with a real world case study for a large client, and the accuracy of machine learning to automate the process end to end.

\end{abstract}

%%
%% The code below is generated by the tool at http://dl.acm.org/ccs.cfm.
%% Please copy and paste the code instead of the example below.
%%


%%
%% Keywords. The author(s) should pick words that accurately describe
%% the work being presented. Separate the keywords with commas.

%%
%% This command processes the author and affiliation and title
%% information and builds the first part of the formatted document.
\maketitle

\section{Introduction}
\label{sec:intro}

Offensive security as a methodology provides services to clients in order to better secure their networks. Penetration testing is one of the most commonly used methodology for offensive security. Even though penetration testing is not a new concept, it has been usually performed using a factory approach: run a scan, check the results, create a report. Getting actual hackers to try and break into a network, application, or device - it is always been difficult to get a genuine, thoughtful, realistic penetration test, and that is what offensive security brought to the table.

Soon, companies began to benefit from offensive security approach by discovering new vulnerabilities that really mattered - nothing encourages action faster than demonstrating how an organization's vulnerability could result in a breach. There is something tangible about penetration testing that is not seen in other methods of vulnerability discovery. Vulnerability scanning and compliance checks provide potential attack paths, theories of how a threat actor \emph{might} use a \emph{potential} vulnerability to gain access. Solid penetration testing results in demonstrable risk, provable methods of gaining real access. 

At the same time, finding new vulnerabilities for clients can be challenging to an organization, based on the sheer number of vulnerabilities that they are already aware of. Companies began asking for help with their vulnerability backlog, which often consisted of millions of findings. Vulnerability scanners such as Qualys \cite{qualys} and Tenable \cite{tenable} pride themselves on the number of vulnerabilities they can detect, however unlike penetration testing results, these vulnerabilities are often not demonstrable. It is hard to tell from scan results, and CVSS scores \cite{cvss}, which vulnerabilities represent real risk, and which are more theoretical \cite{9382369, spring2018towards}. For example, exploitation of a critical finding from a vulnerability scanner may require a threat actor to research and create exploit code based on reverse engineering, while a less critical vulnerability may have already been exploited in the wild, and have easy-to-use exploit modules available to download.

This phenomenon is where the offensive-minded approach is able to help move vulnerability management forward. As threat actors, offensive security practitioners, much more focus on ease and predictability of exploitation for vulnerability analysis.
%It occurred to the offensive security practitioners that, as threat actors, analysis of vulnerabilities is much more focused on ease and predictability of exploitation. 
If there are ready-made exploits available, they would favor those vulnerabilities much more than those that required new code to be written. Indeed, it is rare for penetration testers to create exploit code during a test; this usually occurs based on the tester's specific background and interests, or when the tester finds the same opportunity for exploitation across several tests. Even when a security researcher or penetration tester puts their minds to creating an exploit, some vulnerabilities that are theoretical - those based on unpredictable memory or stack corruption - are not actually paths to code execution. It takes a fair amount of time and creativity to create reliable exploits from the pinhole provided by some vulnerabilities. This observation led to the vulnerabilities that are the most "popular" among threat actors, based on the prevalence of exploit proof-of-concept code. In this paper, we propose a novel approach to rank vulnerabilities based on this offensive security mindset. More specifically, our contributions are as follows:

\begin{itemize}
    \item We introduce three new metrics to capture what penetration testers seek out in their offensive security practices. We call these metrics \emph{weaponized exploit (wx)}, \emph{utility} and \emph{opportune}. 
    \item We propose a stack-ranked scoring mechanism that allows security practitioners to be able to focus on the vulnerabilities that matter the most.
    \item We propose and evaluate a machine learning based approach for automation of the proposed metrics.
    \item We present a case study to showcase how a real-world customer is taking advantage of this approach in comparison to a CVSS-based approach.
\end{itemize}

\section{Background}
\label{sec:background}

% Things to talk about:

% \begin{itemize}
%     \item Why prioritization is needed?
%     \item Current SOTA: CVSS as a prioritization methodology
%     \item Other related prioritization methodology (SVCC-Jono etc.)
% \end{itemize}

\begin{table}[t]
%\scriptsize
\begin{center}
  %\begin{tabular}{ | l | l | m{11cm} |}
  % \vspace{-1mm}
  \begin{tabular}{ | p{3cm} | p{5cm} |}
    \hline
    {\bf Metric} & {\bf Values} \\ \hline \hline
    Attack Vector (AV) & Network (N) | Adjacent (A) | Local (L) | Physical (P)  \\ \hline
    Access Complexity (AC) & Low (L) | High (H)  \\ \hline
    Privileges Required (PR) & None (N) | Low (L) | High (H) \\ \hline
    User Interaction (UI) & None (N) | Required (R) \\ \hline
    Scope (S) & Unchanged (U) | Changed (C) \\ \hline
    Confidentiality (C) & None (N) | Low (L) | High (H) \\ \hline
    Integrity (I) & None (N) | Low (L) | High (H) \\ \hline
    Availability (A) & None (N) | Low (L) | High (H) \\ \hline
    Score:Category Map & $0$: None, $0.1-3.9$: Low, $4.0-6.9$: Medium, $7.0-8.9$: High, $9.0-10.0$: Critical \\ \hline
  \end{tabular}
  \caption{CVSS v3.1 Base Score Metrics}
  \label{table:cvss_metrics}
\end{center}
\vspace{-10mm}
\end{table}

Common Vulnerability Scoring System (CVSS) is a de facto methodology when it comes to vulnerability prioritization \cite{cvss_spec, cvss_calc}. CVSS consists of three main components: Base Score Metrics (BSM), Temporal Score Metrics (TSM) and Environmental Score Metrics (ESM). BSM captures two important aspects: exploitability and the impacts of a vulnerability. Table \ref{table:cvss_metrics} shows the different metrics in CVSS-BSM along with corresponding values for each metric. CVSS has a formula to calculate a score ranges from 0-10 based on the selected values. Based on the score, CVSS also prescribe a way of categorizing the severity of a vulnerability as: None, Low, Medium, High and Critical. 

There are multiple scoring providers for vulnerabilities. National Vulnerability Database (NVD) is one of the widely used scoring provider. NVD analysts examine each and every vulnerability and assess them in terms CVSS-BSM. It is a common practice to leave TSM and ESM to the end users, since the environmental and temporal aspects differ from organization to organization and time to time. It is a common practice for organizations to use BSM only for prioritization since they do not have the resources to tackle the vast number of vulnerabilities by themselves. 

% \begin{figure}[t]
%   \begin{center}
%     \includegraphics[width=1\columnwidth,keepaspectratio]{images/trends.pdf}
%     %\vspace{-3mm}
%     \caption{Number of reported vulnerabilities over the years.}
%     %\vspace{-6mm}
%     \label{fig:trends}
%   \end{center}
% \end{figure}

% \begin{figure}[t]
%   \begin{center}
%     \includegraphics[width=1\columnwidth,keepaspectratio]{images/cves_exploited.png}
%     %\vspace{-3mm}
%     \caption{Timeline of some of the infamous vulnerabilities that are exploited.}
%     %\vspace{-6mm}
%     \label{fig:trends}
%   \end{center}
% \end{figure}

Even though CVSS is great to start with, it is primarily designed to facilitate communication between researcher and software owner in order to perform responsible disclosure. There are two main problems with CVSS that limits its use for vulnerability prioritization. First, CVSS-BSM does not take any current weaponization of the vulnerability into account. One may argue that this is taken care of by the TSM, specifically by the metric \emph{Exploit Code Maturity}. However, the reality is that TSM can only downgrade the score and does not provide the capability needed to track the weaponization of vulnerabilities in wild, and increase the score thereafter. 

%To motivate the problem further, there are currently more than 180K vulnerabilities in NVD and there is a growing number of CVEs reported each year \cite{cvedetails}. Even though NVD only provides BSM, they have shortage of staffs to be able to cope with the number of CVEs. At a given time, there are usually 100s of vulnerabilities, waiting to be scored in official NVD website \cite{nvd_search}. To mitigate the problem, there are AI methodologies to auto assign CVSS metrics for a given vulnerability description \cite{9582205, spanos2018multi, le2019automated, han2017learning, spanos2017assessment}.

Other methodologies have been proposed for vulnerability prioritization, for example, Stakeholder-specific Vulnerability Categorization (SSVC) \cite{spring2020ssvc}. SSVC uses decision trees to make prioritization decisions with a particular focus on stakeholders. The approach aims to accommodate the diversity of the stakeholders involved in the vulnerability management process. The authors consider SSVC as a replacement for CVSS, but SSVC has not been widely adopted in industry. Hence, a feed with vulnerabilities along with SSVC assignments are not available for comparison and require substantial manual efforts to be constructed. 

%In another approach, VULCON \cite{farris2018vulcon} optimizes for time-to-vulnerability remediation and total vulnerability exposure given the number of discovered vulnerabilities, criticality of assets and availability of personnel resources. The goal of VULCUN is to optimize the effort of the security analyst.

Past and future threat activity is an essential component in prioritizing vulnerabilities. Exploit Prediction Scoring System (EPSS) \cite{jacobs2019exploit} assigns a score between 0-100\% indicating how likely a vulnerability can be exploited within the first twelve months after public disclosure. Although this will be invaluable for vulnerabilities that have not been exploited yet, it is still a probability that security practitioners need to account for prioritization and can be used in addition to our approach for future exploitability. The Exploitation component of EPSS is comparable to the threat component in Vulnerability Priority Rating (VPR) \cite{vpr}. VPR reflects the current threat landscape, with values ranging from 0.1 - 10. Tenable Lumin \cite{lumin} uses VPR in combination with an underweighted CVSS score for cyber risk assessment. Lumin and VPR are a propriety tool and methodology, requires human experts in the loop for full coverage, and hence we are not able to reproduce and compare with our approach. 

% Machine learning also plays a vital role in prioritizing vulnerabilities. VPR and EPSS are based on machine learning models and classifier algorithms. The proprietary data used by VPR makes it difficult to customize and adopt. While EPSS provides a flexible implementation for updating the classifier with more data and is based on multiple open datasets, it fails to consider the context of a vulnerability in its prioritization computation. It also lacks the level of explainability for the results required by enterprises using the cloud.

Some other approaches have attempted to capture vulnerability attributes inherent to particular industry domains for vulnerability ranking, such as in robotics with the Robot Vulnerability Scoring System (RVSS) \cite{vilches2018towards}, Internet of Things \cite{anand2021ivqfiot} and medical devices \cite{coley2019rubric}. For instance, RVSS introduced robot safety and environment components and focused on robot-specific attack vectors. These domain-specific methodologies are not widely applicable and thus have limitations with adoption.

%However, these methodologies have not been widely adopted and there is no evidence that companies getting benefit out of them. 

In a nutshell, our methodology differs from previous work in multiple ways. First, our approach takes into account both the current threat landscape (wx) and the future threat landscape (utility and opportune). Second, our approach is fully automated with highly accurate machine learning models, giving 100\% coverage and scores for all the existing and new vulnerabilities.

\section{Methodology}
\label{sec:proposal}

\begin{table*}[t]
\footnotesize
\begin{tabular}{|p{2.2cm}|p{9.2cm}|p{1.2cm}|p{1cm}|p{1cm}|p{1.2cm}|}
\hline
CVE Id & Description & CVSS 3.1 & WX & Utility & Opportune \\ \hline \hline
CVE-2017-0143 & The SMBv1 server in Microsoft Windows Vista SP2; Windows Server 2008 SP2 and R2 SP1; Windows 7 SP1; Windows 8.1; Windows Server 2012 Gold and R2; Windows RT 8.1; and Windows 10 Gold, 1511, and 1607; and Windows Server 2016 allows remote attackers to execute arbitrary code via crafted packets, aka "Windows SMB Remote Code Execution Vulnerability."
&   8.1 (High)   &  26  &    2     &     0      \\ \hline
CVE-2019-11324 &    The urllib3 library before 1.24.2 for Python mishandles certain cases where the desired set of CA certificates is different from the OS store of CA certificates, which results in SSL connections succeeding in situations where a verification failure is the correct outcome. This is related to use of the ssl\_context, ca\_certs, or ca\_certs\_dir argument         &    7.5 (High)      & 2   &    0     &     0      \\ \hline
CVE-2020-27256 & In SOOIL Developments Co., Ltd Diabecare RS, AnyDana-i and AnyDana-A, a hard-coded physician PIN in the physician menu of the insulin pump allows attackers with physical access to change insulin therapy settings. & 6.8 (Medium) & 0 & 2 & 1 \\ \hline
\end{tabular}
\caption{CVE examples with scores and categories}
\label{table:examples}
\vspace{-9mm}
\end{table*}

% Things to talk about:

% \begin{itemize}
%     \item how to capture hacker's point of view? Hacker mindset
%     \item Introduction of wx, utility and opportune
%     \item Automation problem... ML comes to rescue
% \end{itemize}

Taking offensive security as an inspiration, to capture the vulnerabilities that could result in a breach, we propose three additional metrics in addition to the CVSS. These metrics are explained next.

\subsection{Weaponized Exploit (WX)}

One of the very first vulnerabilities that organization would like to focus on fixing are the ones with actual exploit paths. This is also very much what penetration testers use for breaching into the systems and applications. We calculate wx by counting exploit references (as a text) on sites such as ExploitDB, Metasploit, and Github. This is an unbounded count which gives an idea about how widely an exploit is available for a given vulnerability. It also implicitly tells organization that access to the exploitation is quite easy as the score increases. 

\subsection{Utility}

Utility measures how useful the vulnerability appeared in accomplishing the goals of the attacker. This is a useful metric to capture both the current threat and future threat (if not exploited yet). We measure utility in three categories 0, 1 and 2. If utility category is 0, it means that it is not useful for attackers at all to exploit. If utility category is 1, it means that attackers see an attack chaining capability - for example access to a component in a limited way and search/find another vulnerability to move laterally. Lastly if the utility category is 2, it means that actions on objectives for attackers, i.e. these are the top vulnerabilities that the attackers search for. For example remote code exploitation vulnerabilities are usually very attractive to attackers. 

\subsection{Opportune}

Opportune is a measure to more easily find vulnerabilities that may not even require exploit code - for example, default password findings. We capture opportune in two categories: 0 means no opportune, 1 means opportune exists. Vulnerabilities with opportune scores of 1 are the ones that do not require any code to exploit so very attractive to hackers.

Table \ref{table:examples} gives examples of vulnerabilities with CVSS, wx, utility and opportune scores. As can be seen from the Table, only relying on CVSS will not enable organizations to even call these highly exploitable and valuable (from attacker's perspective) vulnerabilities as \emph{Critical} but rather \emph{High} or \emph{Medium}, which hinder organizations' abilities to fastly remediate these highly critical vulnerabilities.

\subsection{Scoring}

In order to be able to prioritize vulnerabilities, we generate one score, which is calculated based on the Equation \ref{eq:scoring}. Our score takes into account both CVSS and the new metrics we introduce. Hence it combines the likelihood of exploitability, impact, the current and future threat landscape. It also includes environmental factors for an asset which includes networking aspect (public vs. private) and criticality - with different weights.
\vspace{-3mm}

\begin{equation}
\label{eq:scoring}
(CVSS + WX) \times (utility + 1) \times (opportune  + 1) \times (Environmental Factors)\\
\end{equation}

This threat score allows us to be able to generate one score that combines wx, utility, opportune and CVSS score itself. Instead of using categorical Critical, High, Medium or Low approach for prioritization, or a bounded formula as prescribed by CVSS, this formula allows us to stack rank the vulnerabilities and focus on the ones that matter the most first. 

\subsection{Automation}

One important aspect of vulnerability prioritization is the scalability. In our case, both CVSS and wx are automated, one provided by NVD and the other is an automated script, crawling different websites and generate the count. On the other hand, utility and opportune categories are manually assigned by Subject Matter Experts (SMEs). Given the size of the vulnerabilities that need to be assessed, it is not feasible for SMEs to go through all of these vulnerabilities and assign utility and opportune scores to each one of them. 

In order to overcome this scalability problem, we utilize machine learning to help us automate the scoring of utility and opportune scores. Our SMEs curated a set of vulnerabilities where they assign utility and opportune scores and we use these as training data to build a machine learning model to be able to auto-assign scores to vulnerabilities. SMEs periodically evaluates new vulnerabilities and hence contributing to obtain better accuracy. Next section discuss the methodology and accuracy of this approach.
\section{Evaluation}
\label{sec:experiments}

% Things to talk about:

% \begin{itemize}
%     \item ML accuracy and discussion
%     \item Demonstration of benefit of wx, utility, opportune over CVSS
%     \item Customer surveys? 
% \end{itemize}

We evaluate our approach in two ways. One to examine the accuracy of machine learning model for end-to-end automation and the second to examine the effectiveness of our approach for a real world client as opposed to using only CVSS for prioritization.

\vspace{-3mm}

\subsection{Machine Learning Accuracy}

\begin{figure*}
    \centering
    \begin{minipage}{0.5\textwidth}
        \centering
        \includegraphics[width=0.9\textwidth]{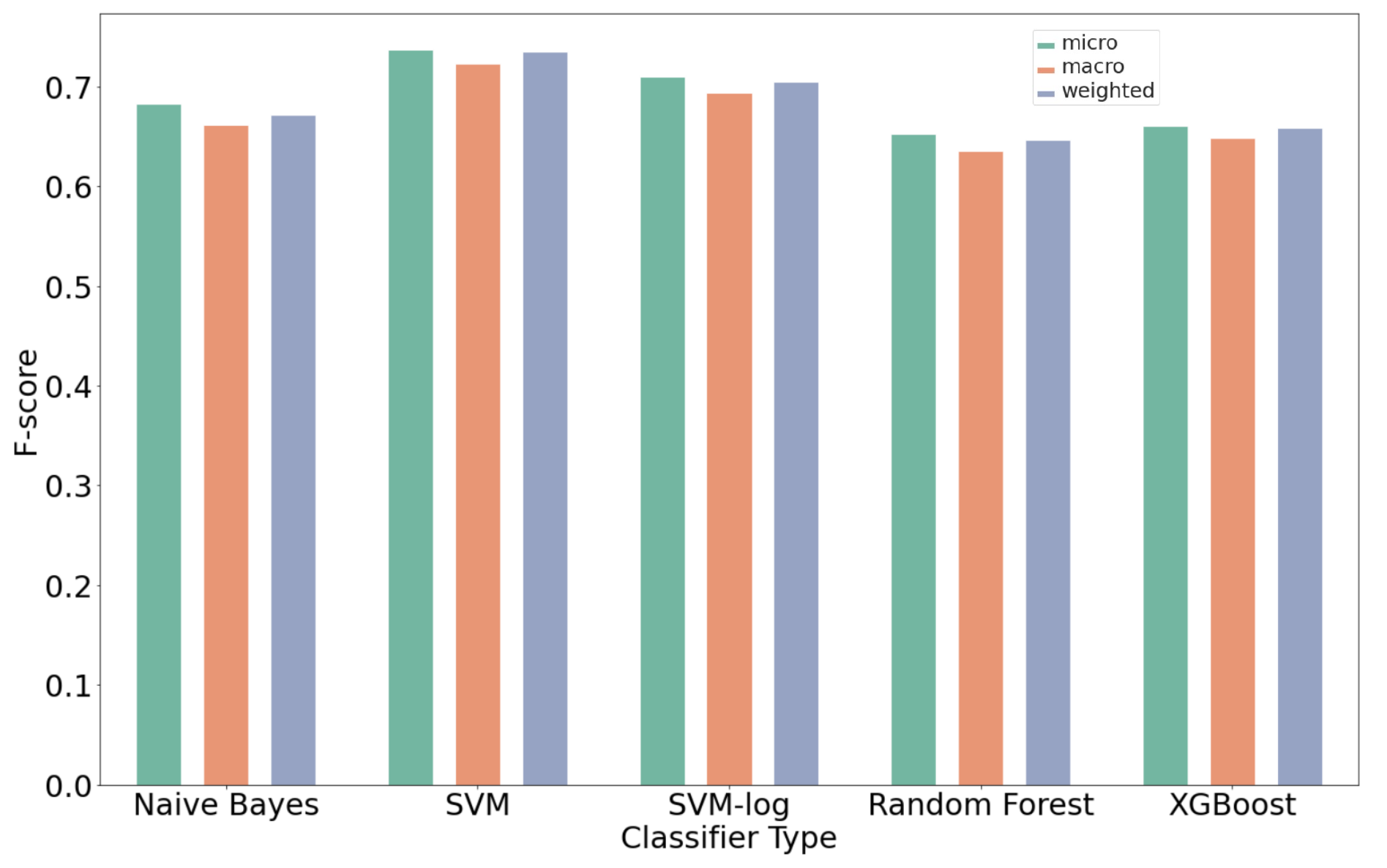} % first figure itself
        \caption{Utility category prediction accuracy.}
        \label{fig:utility}
    \end{minipage}\hfill
    \begin{minipage}{0.5\textwidth}
        \centering
        \includegraphics[width=0.9\textwidth]{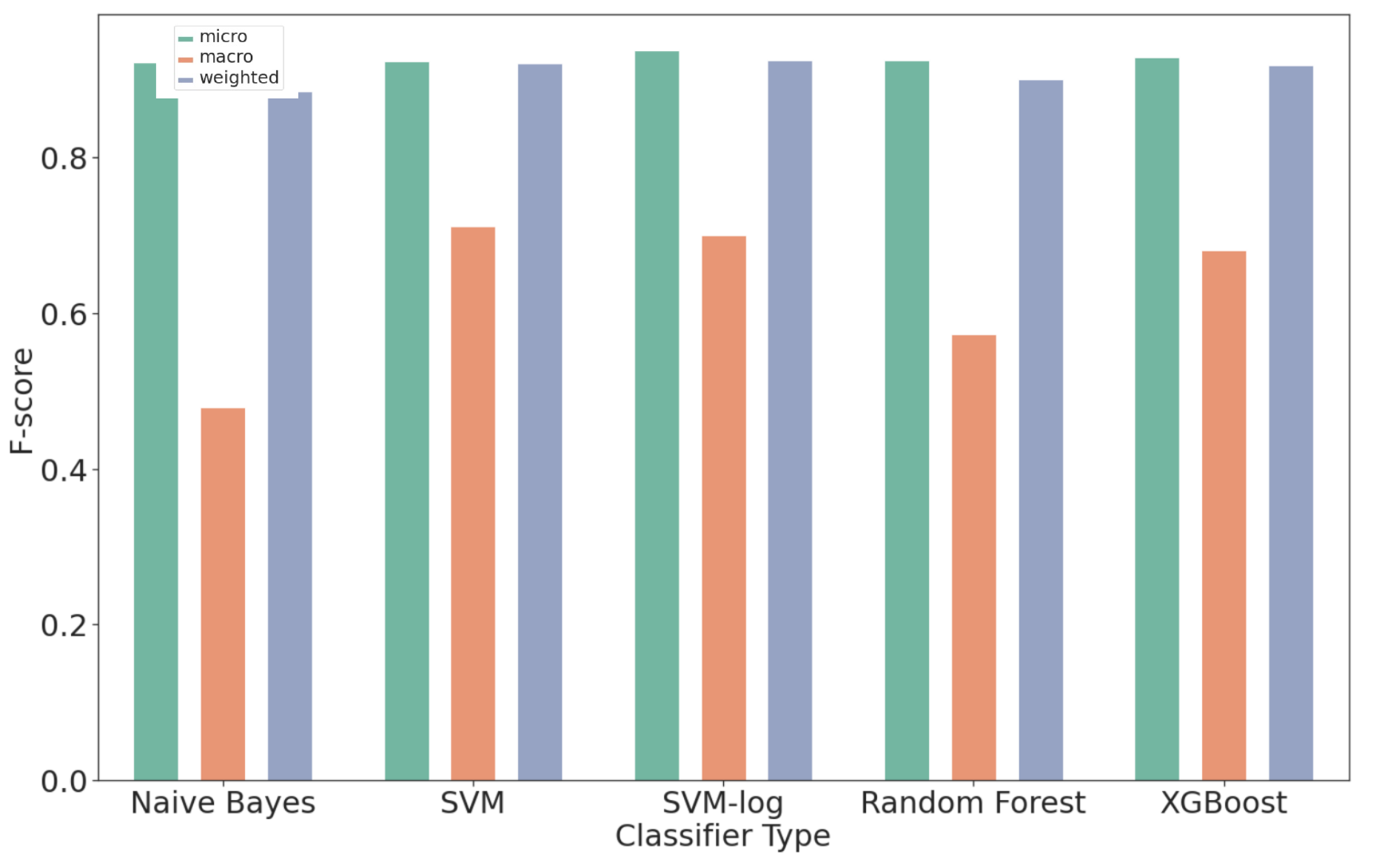} % second figure itself
        \caption{Opportune category prediction accuracy.}
        \label{fig:opportune}
    \end{minipage}
\end{figure*}

\begin{figure*}[t]
  \begin{center}
    \includegraphics[width=1.68\columnwidth,keepaspectratio]{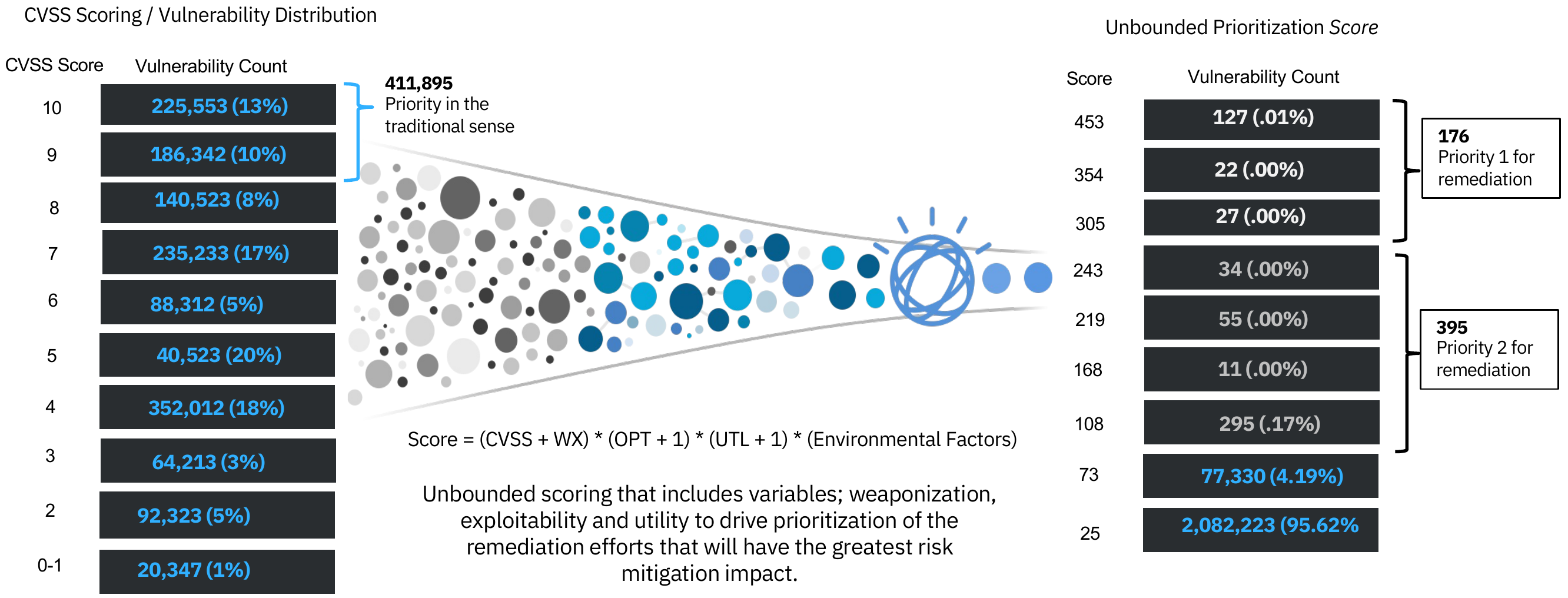}
    %\vspace{-3mm}
    \caption{Prioritization of vulnerabilities according to threat score instead of CVSS.}
    %\vspace{-6mm}
    \label{fig:outcome}
  \end{center}
  \vspace{-5mm}
\end{figure*}

% \begin{figure}[t]
%   \begin{center}
%     \includegraphics[width=1\columnwidth,keepaspectratio]{images/utility.pdf}
%     %\vspace{-3mm}
%     \caption{Utility category prediction accuracy.}
%     %\vspace{-6mm}
%     \label{fig:utility}
%   \end{center}
% \end{figure}

% \begin{figure}[t]
%   \begin{center}
%     \includegraphics[width=1\columnwidth,keepaspectratio]{images/opportune.pdf}
%     %\vspace{-3mm}
%     \caption{Opportune category prediction accuracy.}
%     %\vspace{-6mm}
%     \label{fig:opportune}
%   \end{center}
% \end{figure}

In this section, we evaluate the accuracy of using machine learning for utility and opportune prediction. In order to evaluate the accuracy, we curate a list of vulnerabilities for assignment of utility and opportune categories. This has been done by SMEs who are expert in vulnerability management. In total, we curate 667 vulnerabilities where we have the utility and opportune categories. For utility, 42\% of the data has a category of 0, 32\% of data has a category of 1 and 26\% of the data has category of 2. For opportune, 92\% of the data has an opportune score of 0, and 8\% of the data has an opportune score of 1. Note the unbalance nature of the opportune data. At the same time, this is very much expected as opportune score of 1 are really rare. 

Next we divided 80\% of the data randomly as training and the remaining 20\% as the testing sets. The input to the machine learning model is the description of the vulnerability which are retrieved from NVD. Based on the description we generate tf-idf scores for each one of the description as an input to the model and the model captures the category of the utility and opportune scores. We train two separate models for utility and opportune prediction. In order to evaluate the accuracy of the models we use F-scores. Since the output is multi class we use micro, macro and weighted averages to evaluate the accuracy \cite{fscore}.

Figure \ref{fig:utility} shows the accuracy of prediction for utility categories. With machine learning we achieve F-scores of 74\% (micro), 72\% (macro) and 74\% (weighted) using Support Vector Machine (SVM). Figure \ref{fig:opportune} shows the accuracy of prediction for opportune categories. With machine learning we achieve F-scores of 92\% (micro), 71\% (macro) and 92\% (weighted) using SVM, and 92\% (micro), 68\% (macro), 91\% (weighted) using XGBoost. One reason SVM mostly performs better than other machine learning algorithms might be the fact we have a small dataset for training. XGBoost and decision tree models usually require more data to perform better. 

\vspace{-3mm}

\subsection{Real World Client Use Case}

A case study was performed in order to validate the usefulness of our approach. Figure \ref{fig:outcome} shows the outcome of the study. On the left side of the illustration, the number of vulnerabilities within the environment are listed in order of CVSS v3.1 score, descending from 10 to 1. Commonly, CVSS v3.1 scores of 9-10 are considered "Critical" by organizations, so in this depiction there are 411,895 vulnerabilities that require immediate patching. Based on our experiences with large organizations, patching this number of vulnerabilities within a short time frame is not feasible. Audit teams often proscribe that Critical vulnerabilities be remediated within 7-10 days, with 30 days as the upper limit from our experiences. What is lacking in the example on the left is any indication of which vulnerabilities are being used by real world adversaries and have known weaponization paths. In order to begin reducing these vulnerabilities, we need to know which of these Critical vulnerabilities creates the most risk and focus early remediation efforts accordingly.

On the right, we list the breakdown of vulnerabilities based on the unbounded scoring methods described in this paper. The number of vulnerabilities at the top is considerably reduced, showing that indeed there are heavily targeted vulnerabilities that need to be addressed first. In addition, the different scores provide a contrast between vulnerabilities, providing a sense of how much more risk each vulnerability creates in relation to other similar vulnerabilities. The intent here is to always be able to supply the most impactful next steps to the remediation teams to facilitate an ongoing risk reduction program. There will always be a smaller set of "highest risk vulnerabilities" that is consumable by the remediation teams, regardless of how many are reduced. In this way, we see Vulnerability Management as an ongoing pursuit, not a point-in-time activity. 

%The ability to stack-rank large numbers of vulnerabilities by risk fosters an achievable risk reduction program, both by the teams responsible, and by risk teams that need more granular and achievable goals.  

% \input{discussion}
\section{Conclusion and Future Work}
\label{sec:conclusion}

This paper demonstrates the limitation of CVSS as a prioritization methodology and propose a novel approach that is inspired by offensive security. With the help of machine learning for automation, we are able to rank vulnerabilities and demonstrate the value of the approach with a client use case. To study the problem further, we will focus on increasing the accuracy of the machine learning part and additionally bring business context into the account (e.g. criticality) for better risk-based prioritization. 

% \clearpage
%%
%% The next two lines define the bibliography style to be used, and
%% the bibliography file.
\bibliographystyle{ACM-Reference-Format}
\bibliography{paper}

\end{document}